\documentclass[preprint,authoryear,12pt]{elsarticle}
   \pdfoutput=1
   \usepackage[cp1250]{inputenc}
   \usepackage{xspace}
   \usepackage{graphicx}
   \usepackage{color}
   \usepackage{crayola}
   \usepackage{amssymb}

\journal{Scientometrics}

\begin{document}

\begin{frontmatter}

\title{On bibliographic networks
}
\author{Vladimir Batagelj\corref{cor1}
}
 \ead{vladimir.batagelj@fmf.uni-lj.si}
 \ead[url]{http://pajek.imfm.si}
 \cortext[cor1]{University of Ljubljana, FMF, Department of Mathematics,
   Jadranska 19, 1000 Ljubljana, Slovenia}
\author{Monika Cerin\v sek\corref{cor2}
}
 \ead{monika@hruska.si}
 \cortext[cor2]{Hru\v ska d.o.o., Kajuhova ulica 90, 1000 Ljubljana, Slovenia}
 \address{University of Ljubljana}




\begin{abstract}
In the paper we show that the bibliographic data can be transformed into a
collection of compatible networks. Using network multiplication different
interesting derived networks can be obtained. In defining them an appropriate
normalization should be considered. The proposed approach can be applied also
to other collections of compatible networks. We also discuss the question
when the multiplication of sparse networks preserves sparseness.
The proposed approaches are illustrated with analyses
of collection of networks on the topic "social network" obtained from the
Web of Science.
\end{abstract}


\end{frontmatter}


\newcommand{\Pajek}{\texttt{\textbf{Pajek}}\xspace}
\newcommand{\WoSPajek}{\texttt{\textbf{WoS2Pajek}}\xspace}
\newcommand{\keyw}[1]{\emph{#1}}
\newcommand{\important}[1]{\textcolor{NavyBlue}{#1}}
\newcommand{\RR}{\Bbb{R}}
\newcommand{\NN}{\Bbb{N}}
\newcommand{\ZZ}{\Bbb{Z}}
\newcommand{\QQ}{\Bbb{Q}}
\newcommand{\network}[1]{\mathcal{#1}}
\newcommand{\vertices}[1]{\mathcal{#1}}
\newcommand{\edges}[1]{\mathcal{#1}}
\newcommand{\arcs}[1]{\mathcal{#1}}
\newcommand{\Net}{\network{N}}
\newcommand{\argmin}{\mathop{\rm argmin}\nolimits}
\newcommand{\relation}[1]{\textbf{\emph{$\_\!\_$~#1~$\_\!\_$\,}}}
\newcommand{\card}[1]{\mbox{card}(#1)}
\newcommand{\URL}[1]{{\footnotesize\texttt{#1}}}
\newcommand{\tita}[1]{\textit{#1}}      
\newcommand{\cling}{\mathbf{C}}
\newcommand{\unitX}{\mathrm{X}}
\newcommand{\unitY}{\mathrm{Y}}
\newcommand{\unitZ}{\mathrm{Z}}
\newcommand{\outdeg}{\mathrm{outdeg}}
\newcommand{\indeg}{\mathrm{indeg}}
\newcommand{\ato}{\mathrel{:=}}
\newcommand{\unit}{\mathrm{X}}
\newcommand{\Units}{\vertices{U}}
\def\Min{\mathop{\rm Min}\nolimits}
\def\Max{\mathop{\rm Max}\nolimits}
\newcommand{\Graph}{\mathbf{G}}
\newcommand{\tit}[1]{\textit{#1}}      
\newcommand{\diag}{\mbox{diag}}
\newcommand{\href}[2]{\textit{#2}}
\newcommand{\WA}{\mathbf{W\!\!A}}
\newcommand{\AW}{\mathbf{A\!\!W}}
\newcommand{\WK}{\mathbf{W\!K}}
\newcommand{\KW}{\mathbf{K\!W}}
\newcommand{\WC}{\mathbf{W\!C}}
\newcommand{\WJ}{\mathbf{W\!J}}
\newcommand{\Ci}{\mathbf{Ci}}

\graphicspath{{./pics/}}


\section{Introduction}
\label{s0}

A \emph{collaboration network} is usually defined in the following way. The set of network's
\emph{nodes} consists of \emph{authors}. There exists an \emph{edge} (undirected link) between
authors $u$ and $v$ iff they produced a joint \emph{work} (paper, book, report, etc.).
Its \emph{weight} $w(u,v)$ is equal to the number of works to which $u$ and $v$ both
contributed.

In this case a more basic network is a \emph{two-mode} network linking the set of works
with the set of authors. There is an \emph{arc} (directed link) from the work $p$ to the author $u$ iff
$u$ is an author of the work $p$. It is well known \cite{}
that if we represent this two-mode network with a matrix $\WA$ then we can compute the
matrix of the corresponding collaboration network as $\WA^T*\WA$ using matrix multiplication.

The problem with matrices of large networks is that they require in their standard
representation too much computer memory although most of their entries are zero.
For this reason we introduce a 'parallel' operation of network multiplication that
deals only with nonzero elements.

For a given set of works, besides the two-mode network $\WA$ on works $\times$ authors,
we can construct other two-mode networks such as $\WK$ on works $\times$ keywords,
$\WC$ on works $\times$ classifications,  $\WJ$ on works $\times$ journals, etc.
Since these networks have the same first set -- the set of works,
we can obtain from them using multiplication
different \emph{derived} networks. For example $\WA^T*\WK = \mathbf{AK}$ gives us the
two-mode network $\mathbf{AK}$ on authors $\times$ keywords with the weight of the arc $(u,k)$
counting in how many works the author $u$ used the keyword $k$. Additional
derived networks can be produced considering also the one-mode citation network $\Ci$
between works.

In the paper we first show that we can transform any data table into a collection of
corresponding two-mode networks. Afterwards we introduce the network multiplication
and discuss the question when it preserves the sparsity of networks. Since the
networks from the collection are compatible -- they share a common set -- we can
obtain, using multiplication, different derived networks. The main part of the paper
deals with the problem of 'normalization' of the weights in the derived networks
which is illustrated with the case of collaboration networks. The described approach
can be used also for other derived networks. In the last part of the paper some other
derived networks for the case of bibliographic networks are presented.

The introduced concepts are illustrated on the network data set \textbf{\texttt{SN5}}
obtained in 2008 from the Web of Science for a query \texttt{"social network*"}
and expanded with existing descriptions of the most frequent references and the
bibliographies of around 100 social networkers.
Using the program \WoSPajek \citep{WoS2Pajek}
the corresponding collection of network data was produced: the networks
works $\times$ authors, works $\times$ keywords, ..., citation network;
partition of works by publication year, and the DC partition distinguishing
between works with complete description and the cited only works. The sizes
of the sets are as follows:
works $|W|   =  193376$,
works with complete description $|C|   =  7950$,
authors $|A|   =  75930$,
journals $|J|   =  14651$,
keywords $|K|   =  29267$.
The data set was used  for the Viszards session at the
SunBelt XXVIII, January 22 -- 27, 2008, St Pete Beach, Florida. Analyses were made in a program \Pajek (\cite{pajWiki}), a tool for analysis and visualization of large networks.

\section{Two-mode networks and network multiplication}
\label{s1}

\subsection{Two-mode networks from data tables}

A \keyw{data table} ${\cal T}$ is a set of \keyw{records}
${\cal T} = \{ T_k : k \in \vertices{K} \}$, where $\vertices{K}$ is the set of
\keyw{keys}. A record has the form
$ T_k = (k, q_1(k), q_2(k), \ldots, q_r(k)) $
where $q_i(k)$ is the value of the \keyw{property} (attribute)
$\mathbf{q}_i$ for the key $k$.

Suppose that the property $\mathbf{q}$ has the range $2^\vertices{Q}$.
For example, for \cite{Wasserman1994} :\medskip\\
\indent Authors$($SNA$) = \{$ S. Wasserman, K. Faust $\}$,\\
\indent PubYear$($SNA$) = \{$ 1994 $\}$,\\
\indent Keywords$($SNA$) = \{$ network, centrality, matrix, \ldots $\}$,\ \ldots
\medskip\\
or for \cite{Wasserman1994,GenCores,ESNA2,islands,ifcs}:\medskip\\
\begin{center}
\footnotesize
\begin{tabular}{l|l|r|}
work        &  authors                             & year       \\ \hline
\ldots      &                                      &   \\
SNA         &  S. Wasserman, K. Faust              & 1994  \\
GenCores    &  V. Batagelj, M. Zaver\v snik           & 2011  \\
Islands     &  M. Zaver\v snik, V. Batagelj           & 2004  \\
ESNA2       &  W. de Nooy, A. Mrvar, V. Batagelj   & 2012  \\
IFCS09      &  N. Kej\v zar, S. Korenjak, V. Batagelj & 2010  \\
\ldots      &                                      &   \\ \hline
\end{tabular}
\end{center}
\normalsize\medskip

\noindent
Here work is a key, and authors and year are properties.

If $\vertices{Q}$ is finite  we can assign to the property
$\mathbf{q}$ a two-mode network
$\vertices{K} \times \mathbf{q} = (\vertices{K},\vertices{Q},\edges{A},w)$
where $(k,v) \in \edges{A}$ iff $v \in q(k)$, and $w(k,v) = 1$.
Note that the set $\vertices{Q}$ can always be transformed into a finite set
by partitioning it and recoding the values.\bigskip

Single-valued properties can be represented more compactly by a partition.

For data from the \href{http://home.izum.si/izum/ft_baze/wos.asp}{Web of Science}
(Knowledge) we can obtain the corresponding networks using the program
\href{http://pajek.imfm.si/doku.php?id=wos2pajek}{\WoSPajek} \citep{WoS2Pajek}.
Similar programs exist also for other bibliographic data sources/formats: BiB\TeX,
DBPL, IMDB, Zentralblatt Math, and others.

\subsection{Multiplication of networks}

The product of two compatible networks is essentially the network corresponding to
the product of matrices corresponding to the given networks; or in more formal words:

To a simple (no parallel arcs) two-mode \keyw{network}
$\network{N} = (\vertices{I},\vertices{J},\edges{A},w)$;
where $\vertices{I}$ and $\vertices{J}$ are
sets of \keyw{nodes}, $\edges{A}$ is a set of \keyw{arcs}
linking $\vertices{I}$ and $\vertices{J}$, and
$w : \edges{A} \to \RR$  is a \keyw{weight};
we can assign a \keyw{network matrix}
$\mathbf{W} = [ w_{i,j} ]$ with elements: $w_{i,j} = w(i,j)$ for
$(i,j) \in \edges{A}$ and $w_{i,j} = 0$ otherwise.

Given a pair of compatible two-mode networks
$\network{N}_A = (\vertices{I},\vertices{K},\edges{A}_A,w_A)$ and
$\network{N}_B = (\vertices{K},\vertices{J},\edges{A}_B,w_B)$ with
corresponding matrices $\mathbf{A}_{\vertices{I} \times \vertices{K}}$
and $\mathbf{B}_{\vertices{K} \times \vertices{J}}$
we call a \keyw{product of networks} $\network{N}_A$ and $\network{N}_B$ a network
$\network{N}_C = (\vertices{I},\vertices{J},\edges{A}_C,w_C)$,
where $\edges{A}_C = \{ (i,j): i \in \vertices{I}, j \in \vertices{J}, c_{i,j} \ne 0 \}$
and $w_C(i,j) = c_{i,j}$ for $(i,j) \in \edges{A}_C$. The product matrix
 $\mathbf{C} = [ c_{i,j} ]_{\vertices{I} \times \vertices{J}} = \mathbf{A} * \mathbf{B}$
is defined in the standard way
\begin{equation}
c_{i,j} = \sum_{k \in \vertices{K}} a_{i,k} \cdot b_{k,j} \label{prod}
\end{equation}
In some applications we have to consider the product on other semirings than the
standard $(\RR,+,\cdot,0,1)$ \citep{semi}.

In the case when $\vertices{I} = \vertices{K} = \vertices{J}$ we are dealing with ordinary one-mode
networks with square matrices.

\begin{figure}[h]
\centerline{\includegraphics[width=90mm,bb=30 10 310 220,clip=]{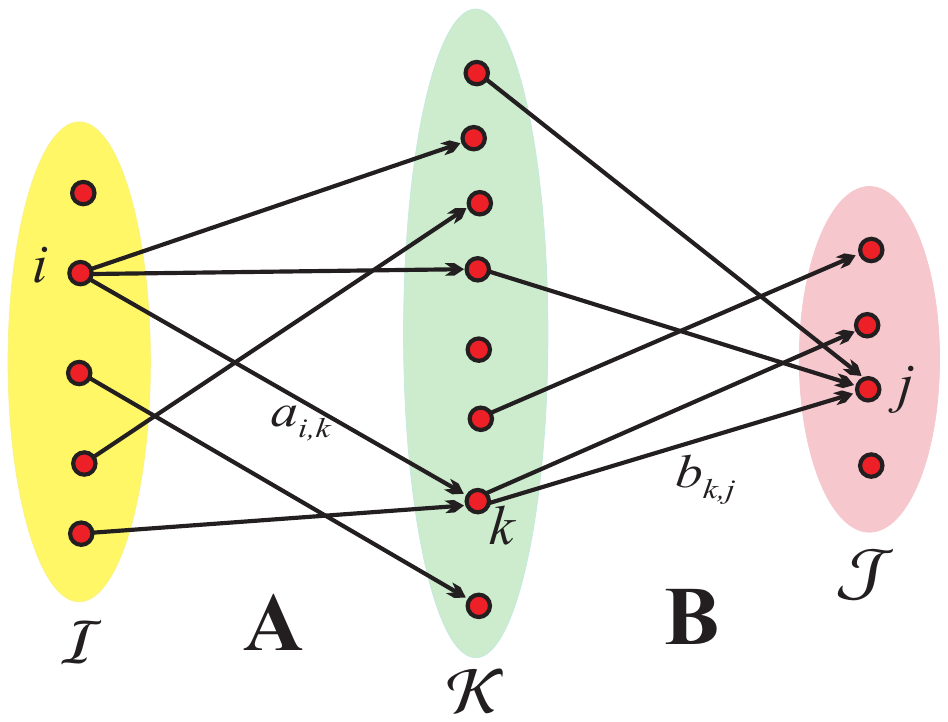}}
\caption{Network multiplication}\label{mult}
\end{figure}

Note that in the expression (\ref{prod}) to the value $c_{i,j}$ contribute only the terms
$a_{i,k} \cdot b_{k,j}$ in which both factors $a_{i,k}$ and $b_{k,j}$ are nonzero.
For $N_A(i)\cup N^-_B(j) \ne \emptyset$ we have
\[ c_{i,j} = \sum_{k \in N_A(i)\cup N^-_B(j)} a_{i,k} \cdot b_{k,j} \]
where $N_A(i)$ are the \keyw{successors} of node $i$ in network $\network{N}_A$
and $N^-_B(j)$ are the \keyw{predecessors} of node $j$ in network $\network{N}_B$.

Therefore, if all weights in networks $\network{N}_A$ and $\network{N}_B$ are equal to 1
then the product $a_{i,k} \cdot b_{k,j} \in \{0,1\}$ and the
value of $c_{i,j}$ counts the number of ways we can go from $i \in \vertices{I}$
to $j \in \vertices{J}$ passing through $\vertices{K}$.

The standard matrix multiplication has the complexity
$O(|\vertices{I}|\cdot |\vertices{K}|\cdot |\vertices{J}|)$ -- it is
too slow to be used for large networks.
Most of large networks are \emph{sparse}  -- their matrices contain much
more zero elements than nonzero elements.
For sparse large networks we can multiply much faster considering only
nonzero elements.

\begin{tabbing}
xxx\=xxx\=xxx\=xxx\=xxx\kill
\>\textbf{for} $k$ \textbf{in} $\vertices{K}$ \textbf{do} \\
\>\> \textbf{for} $(i,j)$ \textbf{in} $N^-_A(k) \times N_B(k)$ \textbf{do}  \\
\>\>\> \textbf{if} $\exists c_{i,j}$ \textbf{then}
         $c_{i,j} := c_{i,j} + a_{i,k} \cdot b_{k,j}$ \\
\>\>\> \textbf{else} new $c_{i,j} := a_{i,k} \cdot b_{k,j}$
\end{tabbing}

In general the multiplication of large sparse networks is a
'dangerous' operation since the result can
'explode' -- it is not sparse.

From the network multiplication algorithm we see that each intermediate node
$k \in \vertices{K}$ adds to a product network a complete two-mode subgraph
$K_{N^-_A(k),N_B(k)}$ (or, in the case $\vertices{I} = \vertices{J}$, a complete
subgraph $K_{N(k)}$). If both degrees $\deg_A(k)=|N^-_A(k)|$
and $\deg_B(k)=|N_B(k)|$ are large then
already the computation of this complete subgraph has a quadratic (time and space)
complexity -- the result 'explodes'.

It is easy to see that if at least one of the sparse networks $\network{N}_A$ and
$\network{N}_B$ has small maximal degree on $\vertices{K}$ then also the resulting
product network $\network{N}_C$ is sparse.

We shall prove a stronger result that if for the sparse networks $\network{N}_A$ and
$\network{N}_B$ there are in $\vertices{K}$ only some vertices  with large
degree and no one among them with large degree in both networks
then also the resulting product network $\network{N}_C$ is sparse.

Let
\[ d_{min}(k) = \min(\deg_A(k), \deg_B(k)) \quad \mbox{and} \quad
 d_{max}(k) = \max(\deg_A(k), \deg_B(k)) .\]
Then
\[ \deg_A(k) \cdot \deg_B(k) = d_{min}(k) \cdot d_{max}(k) \]
Define also
$\Delta_{min} = \max_{k \in \vertices{K}} d_{min}(k)$
and
\[ \vertices{K}(d) = \{ k \in \vertices{K}: d_{max}(k) \geq d \} \]
Let us denote
$d^* = \argmin_d ( |\vertices{K}(d)| \leq d )$ and $\vertices{K}^* =\vertices{K}(d^*)$.
Then $|\vertices{K}^*| \leq d^*$ and the number of nonzero elements in the product

\[ C
   \leq  \sum_{k \in \vertices{K}} \deg_A(k) \cdot \deg_B(k) =
      \sum_{k \in \vertices{K}} d_{min}(k) \cdot d_{max}(k) \]
\[
   =  \sum_{k \in \vertices{K}^*} d_{min}(k) \cdot d_{max}(k) +
   \sum_{k \in \vertices{K}\setminus \vertices{K}^*} d_{min}(k) \cdot d_{max}(k)  \]
\[
   \leq  \Delta_{min} \cdot \sum_{k \in \vertices{K}^*}  d_{max}(k) +
  d^*  \cdot  \sum_{k \in \vertices{K}\setminus \vertices{K}^*} d_{min}(k)  \]
\[ \leq  d^* \cdot ( \Delta_{min} \cdot \max(|\vertices{I}|,|\vertices{J}|)  +
          \min( |\edges{A}_A| , |\edges{A}_B|) )\]
Therefore:
\begin{quote}
If for the sparse networks $\network{N}_A$ and
$\network{N}_B$ the quantities $\Delta_{min}$ and $d^*$
are small  then also the resulting product network $\network{N}_C$ is sparse.
\end{quote}
That is equivalent to the claimed result.

\section{Collaboration}

\subsection{Co-authorship networks}

Let $\mathbf{\WA}$ be the works $\times$ authors two-mode co-authorship
network; $wa_{pi} \in \{0,1\}$ is describing
the authorship of author $i$ of work $p$. Then for each work $p \in W$~:
\[  \sum_{i \in A} wa_{pi} = \outdeg(p) \]
The $\outdeg(p)$ is equal to the number of authors of work $p$.

Let $\mathbf{N}$ be its normalized version with $n_{pi}$ describing the share of
contribution of author $i$ to work $p$ such that for each work $ p \in W$~:
\[  \sum_{i \in A} n_{pi} \in \{0, 1\} \]
The sum has value 0 for works without authors.

The contributions $n_{pi}$ can be determined by some rules or, assuming that each author
contributed equally to the work, it can be computed from $\mathbf{\WA}$ as
\[ n_{pi} =  \frac{wa_{pi}}{\max(1,\outdeg(p))} . \]
A similar normalization of collaboration links, but with $\outdeg(p)-1$ instead of
$\outdeg(p)$, was proposed already by \cite{mej}. He is interpreting the weight as
a proportion of time spent for the collaboration with each co-author.

\keyw{Row-normalization} $n(\network{N})$ is a network obtained from $\network{N}$
in which the weight of each arc $a$ is divided by the sum of weights of all arcs
having the same initial node as the arc $a$. For binary network
$\mathbf{A}$ on $\vertices{I}\times \vertices{J}$
 \[ n(\mathbf{A}) = \diag\left(\frac{1}{\max(1,\outdeg(i))}\right)_{i \in \vertices{I}} * \mathbf{A} \]
Therefore we can obtain the normalized co-authorship network as
\[ \mathbf{N} = n(\WA) \]

In some sense reverse transformation is the \keyw{binarization} $b(\network{N})$ of
the $\network{N}$: it is the original network in which all weights are set to $1$.
It holds
\[ \WA = b(\mathbf{N}) \]
and if $\mathbf{N}$ was obtained from $\WA$ also $\WA = b(n(\WA))$.

Another useful transformation is the transposition.
\keyw{Transposition} $\network{N}^T$ or $t(\network{N})$
is a network obtained from $\network{N}$
in which to all arcs their direction is reversed. For bibliographic networks
we introduce the abbreviations  $\AW = \WA^T$, $\KW = \WK^T$, etc.

\subsection{The first collaboration network}

A standard way to obtain the collaboration network $\mathbf{Co}$ from the
co-authorship network using network multiplication is
\[ \mathbf{Co} = \AW * \WA \]
From
\[ co_{ij} = \sum_{p \in W} wa_{pi} wa_{pj} = \sum_{p \in N(i)\cap N(j)} 1 \]
we see that
$co_{ij}$ is equal to the number of works that authors $i$ and $j$ wrote together.

The weights in the first collaboration network are symmetric
\[ co_{ij} = \sum_{p \in W} wa_{pi} wa_{pj} = \sum_{p \in W} wa_{pj} wa_{pi} = co_{ji} \]

One can search for authors with most collaborators. Such authors in the set \texttt{SN5} are listed in Table~\ref{Tcoll}. On the other hand Table~\ref{Tfreq} shows the distribution of output degree of authors in the set \texttt{SN5}. Output degree of each author is equal to the number of works he/she co-authored.

\begin{table}[h]
\caption{List of the authors with the largest number of different collaborators in \texttt{SN5}%
  \label{Tcoll}}

\begin{center}
\scriptsize
\begin{tabular}{r|l|r||r|l|r|}
i & author & collaborators & i & author & collaborators \\
\hline
   1 &  Snijders,T    &     77 &    11 &  Rothenberg,R  &     58 \\
   2 &  Krackhardt,D  &     71 &    12 &  Doreian,P     &     56 \\
   3 &  Wasserman,S   &     65 &    13 &  Breiger,R     &     56 \\
   4 &  Ferligoj,A    &     63 &    14 &  Valente,T     &     52 \\
   5 &  Berkman,L     &     63 &    15 &  Butts,C       &     52 \\
   6 &  van Duijn,M   &     63 &    16 &  Goodreau,S    &     52 \\
   7 &  Donovan,D     &     62 &    17 &  Draper,D      &     51 \\
   8 &  Friedman,S    &     60 &    18 &  Batagelj,V    &     51 \\
   9 &  Latkin,C      &     59 &    19 &  Barabasi,A    &     51 \\
  10 &  Faust,K       &     59 &    20 &  Kelly,J       &     50 \\
\hline
\end{tabular}
\end{center}

\end{table}

\begin{table}[h]
\caption{Outdegree distribution in $\mathbf{WA}$(\texttt{SN5})%
  \label{Tfreq}}

\begin{center}
\scriptsize
\begin{tabular}{r|r||r|r|l|}
outdeg & frequency & outdeg & frequency & paper \\
\hline
  1 &   2637  &    12 &      8 & \\
  2 &   2143  &    13 &      4 & \\
  3 &   1333  &    14 &      3 & \\
  4 &    713  &    15 &      2 & \\
  5 &    396  &    21 &      1 & \cite{coreG} \\
  6 &    206  &    22 &      1 & \cite{coreF} \\
  7 &    114  &    23 &      1 & \cite{coreE} \\
  8 &     65  &    26 &      1 & \cite{coreD} \\
  9 &     43  &    41 &      1 & \cite{coreC} \\
 10 &     24  &    42 &      1 & \cite{coreB} \\
 11 &     10  &    48 &      1 & \cite{coreA} \\
\hline
\end{tabular}
\end{center}

\end{table}

The obvious question is: who are the most collaborative authors?
The standard answer is provided by $k$-cores, \cite{GenCores}.

A subset $\vertices{U} \subseteq \vertices{V}$ of nodes determines a \keyw{$k$-core}
$\network{C} = (\vertices{U},\edges{L}|\vertices{U})$ in the network
$\network{N} = (\vertices{V},\edges{L})$ iff for each node $u \in \vertices{U}$ it
holds $\deg_\network{C}(u) \geq k$ and the set $\vertices{U}$ is the maximal such set.
The subset of links $\edges{L}|\vertices{U}$ consists of links from $\edges{L}$
that have both end-nodes in $\vertices{U}$.

\begin{figure}[h]

\centerline{%
\includegraphics[width=0.99\textwidth,bb=20 20 990 720,clip=]{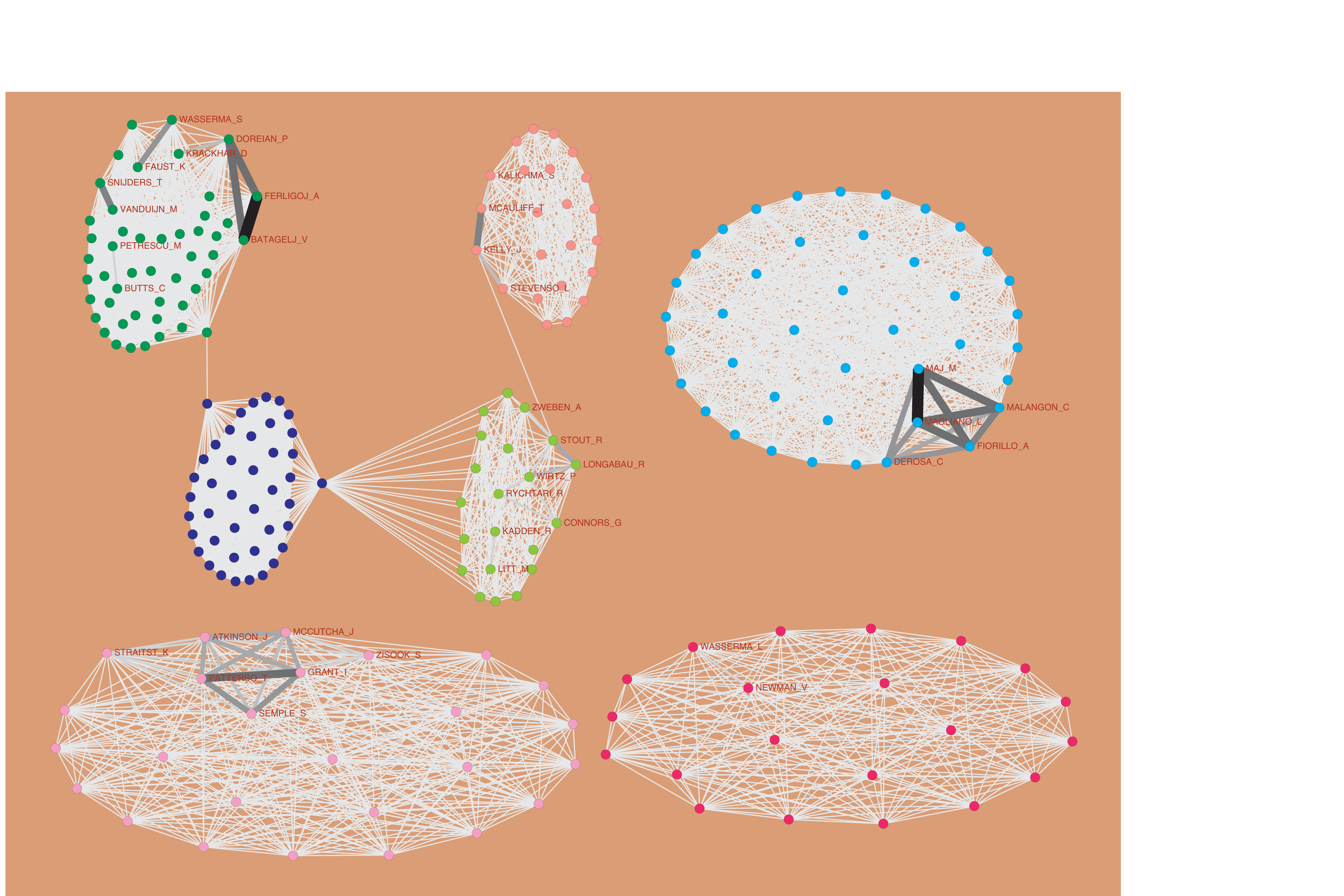}}
\caption{Cores of orders 20--47 in $\mathbf{Co}$(\texttt{SN5})%
  \label{Pcore}}

\end{figure}

In a collaboration network a $k$-core is the largest subnetwork with the property that
each its author wrote a joint work with at least $k$ other authors from the core.

In Figure~\ref{Pcore} the cores of orders 20-47 are presented. From this figure we can see
a serious drawback of directly applying cores for analysis of collaboration networks.
A work with $k$ authors contributes a complete subgraph on $k$ vertices to a collaboration
network. For the bibliographies with works with large number of authors the cores
procedure identifies as the highest level cores the complete subgraphs corresponding
to these works, and not the groups of really the most collaborative authors, as one would
expect.

For the SN5 bibliography the components of the cores of orders 20-47 are induced by the
papers \cite{coreA,coreB,coreC,coreD,coreE,coreF,coreG} that correspond to the works
with the largest number of authors (21-48), see Table~\ref{Tfreq} and \ref{appendix}.
In the picture only the names of authors that are the end-nodes of links with weight
larger than 1 are displayed.

An approach to deal with this problem would be to remove all links with weight 1 (or
up to some other small threshold) and apply cores on the so reduced network.

A better solution is to identify the works with (too) many authors -- very high outdegree
in the network $\mathbf{\WA}$ -- and,  for this analysis, remove them from the network
$\mathbf{\WA}$. We can review the removed works separately.

Yet another approach is to apply on the collaboration network $\mathbf{Co}$ the
$p_S$-cores \citep{GenCores} -- a generalization of the ordinary cores in which
the degree $\deg_\network{C}(u)$ is replaced by the sum of weights of links from
$u$ to other nodes in $\vertices{U}$
\[  p_S(u,\vertices{U}) = \sum_{v \in \vertices{U}} w(u,v) \]

A subset $\vertices{U} \subseteq \vertices{V}$ of nodes determines a \keyw{$p_S$-core} at level $t$
$\network{C} = (\vertices{U},\edges{L}|\vertices{U})$ in the network
$\network{N} = (\vertices{V},\edges{L})$ iff for each node $u \in \vertices{U}$ it
holds $p_S(u,\vertices{U}) \geq t$ and the set $\vertices{U}$ is the maximal such set.

\begin{figure}

\centerline{\includegraphics[height=0.88\textwidth,bb=25 70 985 715,angle=90,clip=]{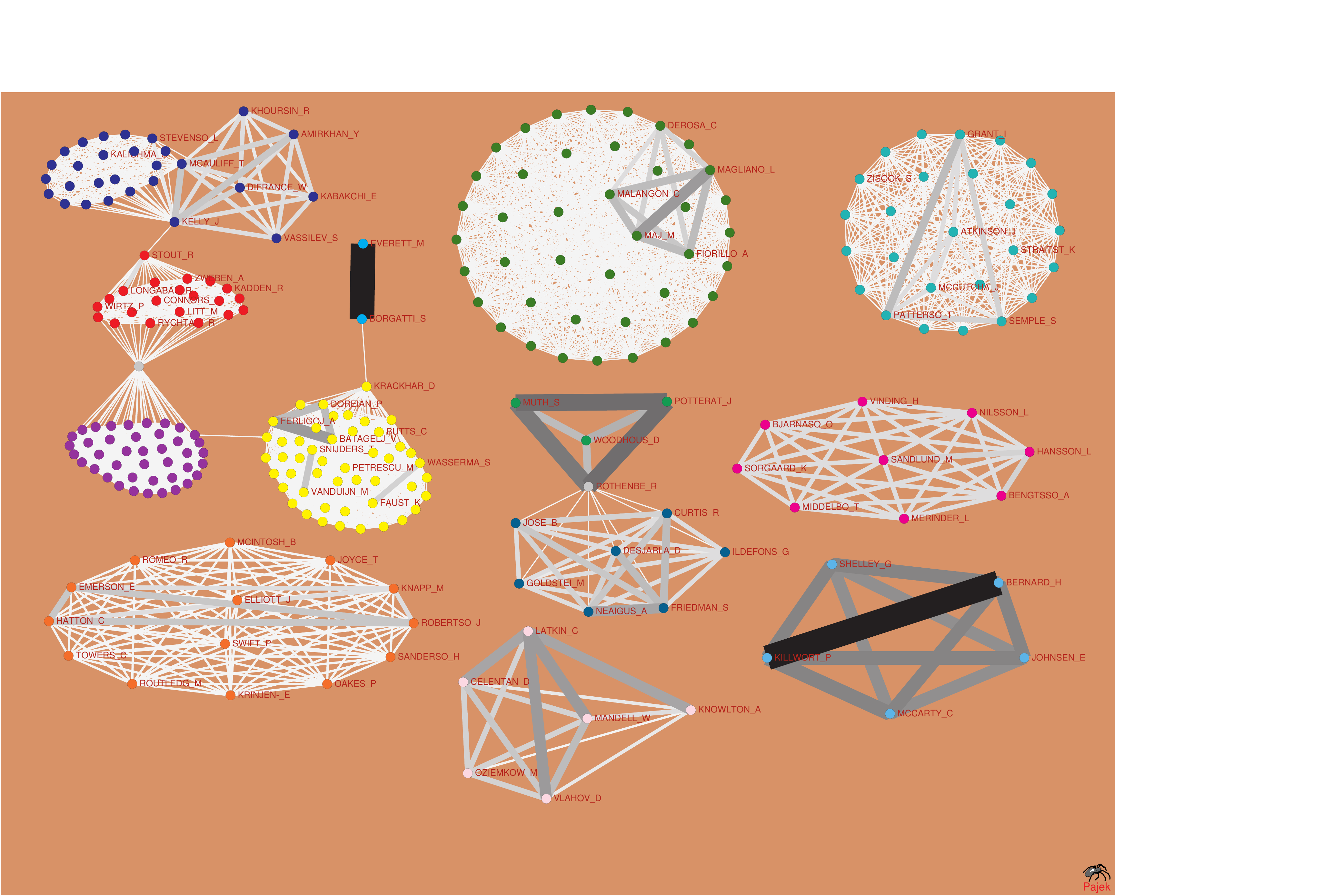}}
\caption{$p_S$-core at level 20 of $\mathbf{Co}$(\texttt{SN5})%
  \label{Pscore}}

\end{figure}

In Figure~\ref{Pscore} the $p_S$-core at level 20 is presented. Each author
belonging to it has at least 20 collaborations with other authors inside the core.

Again in the network \texttt{SN5} the cliques corresponding to papers with the largest number
of authors appear in the $p_S$-core. Besides them we get also  some strongly
collaborating groups such as:
$\{$ S. Borgatti, M. Everett $\}$,
$\{$ H. Bernard, P. Killworth, C. McCarty, E. Johnsen, G. Shelley $\}$,
$\{$ R. Rotenberg, S. Muth, J. Potterat, D. Woodhouse $\}$,
$\{$ L. Magliano, M. Maj, C. Malangon, A. Fiorillo $\}$,
and others.

To neutralize the overrating of the contribution of works with many authors we can
try with alternative definitions of collaboration networks using the normalized
co-authorship network. The structure (graph) of the collaboration network remains
the same, but the weights change.

\subsection{The second collaboration network}
\noindent is defined as

\[ \mathbf{Cn} = \AW*\mathbf{N} \]
The value of the weight $cn_{ij}$
\[ cn_{ij} = \sum_{p \in W} wa_{pi} n_{pj} = \sum_{p \in N(i)\cap N(j)} n_{pj} \]
is equal to the contribution of author $j$ to works, that he/she wrote together with
the author $i$.

In general the entries of matrix $\mathbf{Cn}$ need not to be symmetric
($cn_{ij}=cn_{ji}$). In the case when $n_{pi} =  \frac{wa_{pi}}{\outdeg_{\WA}(p)}$ they are

\[ cn_{ij} = \sum_{p \in N(i)\cap N(j)} n_{pj} =
   \sum_{p \in N(i)\cap N(j)} \frac{1}{\outdeg_{\WA}(p)} = \sum_{p \in N(j)\cap N(i)} n_{pi} =
   cn_{ji} \]

The total contribution of terms $wa_{pi} n_{pj}$ for a work $p$ from the definition
of $cn_{ij}$
\[ \sum_{j \in A} \sum_{j \in A} wa_{pi} n_{pj} =  \sum_{i \in N(p)} \sum_{j \in A} n_{pj} =
   \sum_{i \in N(p)} 1 = \outdeg_{\WA}(p) \]
is equal to the number of authors of the work $p$.

Similary, for an author $i$ the total contribution of entries $cn_{ij}$
\[ \sum_{j \in A} cn_{ij} = \sum_{j \in A} \sum_{p \in W} wa_{pi} n_{pj} =
 \sum_{p \in W} wa_{pi} \sum_{j \in A} n_{pj} = \sum_{p \in W} wa_{pi} = \indeg_{\WA}(i) \]
is equal to the number of works that the author $i$ co-authored; and the
(diagonal) entry
\[ cn_{ii}  = \sum_{p \in N(i)} n_{pi} \]
is equal to the total contribution of author $i$ to his/her works.

\begin{table}
\caption{List of the "best"~authors in SN5}

\begin{center}
\scriptsize
\begin{tabular}{r|l|rrr|}
i & author & $cn_{ii}$ & total & $K_i$ \\
\hline
  1 & Burt,R        &     $ 43.83$ & $53$  &  $0.173$ \\
  2 & Newman,M      &     $ 36.77$ & $60$  &  $0.387$ \\
  3 & Doreian,P     &     $ 34.44$ & $47$  &  $0.267$ \\
  4 & Bonacich,P    &     $ 30.17$ & $41$  &  $0.264$ \\
  5 & Marsden,P     &     $ 29.42$ & $37$  &  $0.205$ \\
  6 & Wellman,B     &     $ 26.87$ & $41$  &  $0.345$ \\
  7 & Leydesdorf,L  &     $ 24.37$ & $35$  &  $0.304$ \\
  8 & White,H       &     $ 23.50$ & $33$  &  $0.288$ \\
  9 & Friedkin,N    &     $ 20.00$ & $23$  &  $0.130$ \\
 10 & Borgatti,S    &     $ 19.20$ & $41$  &  $0.532$ \\
 11 & Everett,M     &     $ 16.92$ & $31$  &  $0.454$ \\
 12 & Litwin,H      &     $ 16.00$ & $21$  &  $0.238$ \\
 13 & Freeman,L     &     $ 15.53$ & $20$  &  $0.223$ \\
 14 & Barabasi,A    &     $ 14.99$ & $35$  &  $0.572$ \\
 15 & Snijders,T    &     $ 14.99$ & $30$  &  $0.500$ \\
 16 & Valente,T     &     $ 14.80$ & $34$  &  $0.565$ \\
 17 & Breiger,R     &     $ 14.44$ & $20$  &  $0.278$ \\
 18 & Skvoretz,J    &     $ 14.43$ & $27$  &  $0.466$ \\
 19 & Krackhardt,D  &     $ 13.65$ & $25$  &  $0.454$ \\
 20 & Carley,K      &     $ 12.93$ & $28$  &  $0.538$ \\
 21 & Pattison,P    &     $ 12.10$ & $27$  &  $0.552$ \\
 22 & Wasserman,S   &     $ 11.72$ & $26$  &  $0.549$ \\
 23 & Berkman,L     &     $ 11.21$ & $30$  &  $0.626$ \\
 24 & Moody,J       &     $ 10.83$ & $15$  &  $0.278$ \\
 25 & Scott,J       &     $ 10.47$ & $15$  &  $0.302$ \\
 26 & Latkin,C      &     $ 10.14$ & $37$  &  $0.726$ \\
 27 & Morris,M      &     $  9.98$ & $20$  &  $0.501$ \\
 28 & Rothenberg,R  &     $  9.82$ & $28$  &  $0.649$ \\
 29 & Kadushin,C    &     $  9.75$ & $11$  &  $0.114$ \\
 30 & Faust,K       &     $  9.72$ & $18$  &  $0.460$ \\
 31 & Batagelj,V    &     $  9.69$ & $20$  &  $0.516$ \\
 32 & Mizruchi,M    &     $  9.67$ & $15$  &  $0.356$ \\
 33 & [Anon]        &     $  9.00$ & $ 9$  &  $0.000$ \\
 34 & Johnson,J     &     $  8.89$ & $21$  &  $0.577$ \\
 35 & Fararo,T      &     $  8.83$ & $16$  &  $0.448$ \\
 36 & Lazega,E      &     $  8.50$ & $12$  &  $0.292$ \\
 37 & Knoke,D       &     $  8.33$ & $11$  &  $0.242$ \\
 38 & Ferligoj,A    &     $  8.19$ & $19$  &  $0.569$ \\
 39 & Brewer,D      &     $  8.03$ & $11$  &  $0.270$ \\
 40 & Klovdahl,A    &     $  7.96$ & $17$  &  $0.532$ \\
 41 & Hammer,M      &     $  7.92$ & $10$  &  $0.208$ \\
 42 & White,D       &     $  7.83$ & $15$  &  $0.478$ \\
 43 & Holme,P       &     $  7.42$ & $14$  &  $0.470$ \\
 44 & Boyd,J        &     $  7.37$ & $13$  &  $0.433$ \\
 45 & Kilduff,M     &     $  7.25$ & $16$  &  $0.547$ \\
 46 & Small,H       &     $  7.00$ & $ 7$  &  $0.000$ \\
 47 & Iacobucci,D   &     $  7.00$ & $12$  &  $0.417$ \\
 48 & Pappi,F       &     $  6.83$ & $10$  &  $0.317$ \\
 49 & Chen,C        &     $  6.78$ & $12$  &  $0.435$ \\
 50 & Seidman,S     &     $  6.75$ & $ 9$  &  $0.250$ \\
 \hline
 \end{tabular}
 \end{center}

 \end{table}

Therefore we can base on the entries of matrix $\mathbf{Cn}$ the
\keyw{self-sufficiency} index
\[ S_i = \frac{cn_{ii}}{\indeg_{\WA}(i)} \]
as the proportion of author's contribution to his/her works and the total
number of works he/she co-authored.

The \keyw{collaborativness} index is complementary to it
\[ K_i = 1 - S_i \]
All $cn_{ij}$ values
\[ \sum_{i \in A} \sum_{j \in A} cn_{ij} = \sum_{i \in A} \indeg_{\WA}(i) = m_{\WA}\]
sum up to the number of all links in the network $\WA$.

In Table~3 the 50 authors with the largest self-contribution $cn_{ii}$ to the topic of
'social network analysis' are presented together with the total number of works on the
topic that an author co-authored, and his/her collaborativness index.

\subsection{The third collaboration network}
\noindent is defined as

\[ \mathbf{Ct} = \mathbf{N}^T * \mathbf{N} \]

The weight $ct_{ij}$ is equal to the total contribution of collaboration of authors
$i$ and $j$ to works.

The total contribution of a complete subgraph corresponding to the authors of a work is 1:

\[ \sum_{i \in A} \sum_{j \in A} n^T_{ip} n_{pj} =
   \sum_{i \in A} n_{pi} \sum_{j \in A} n_{pj} = \sum_{i \in A} n_{pi} \cdot 1 = 1 \]
The weights $ct_{ij}$ are symmetric
\[ ct_{ij} = \sum_{p \in W} n^T_{ip} n_{pj} = \sum_{p \in W} n^T_{jp} n_{pi} = ct_{ji} \]
and the sum
\[ \sum_{j \in A} ct_{ij} = \sum_{j \in A} \sum_{p \in W} n_{pi} n_{pj} =
 \sum_{p \in W} n_{pi} \sum_{j \in A} n_{pj} = \sum_{p \in W} n_{pi}  \]
is equal to the total contribution of author $i$ to works from $W$.

The sum of all weights $ct_{ij}$
\[ \sum_{i \in A} \sum_{j \in A} ct_{ij} = \sum_{i \in A} \sum_{p \in W} n_{pi} =
 \sum_{p \in W} \sum_{i \in A} n_{pi} =  \sum_{p \in W} 1 = |W| \]
is equal to the number of all works.

We can also introduce the \keyw{author's contribution to the field} as

\[ ac_i = \frac{|A|}{|W|} \sum_{p \in W} n_{pi} \]
with the property
\[ \sum_{i \in A} ac_i = |A| \]
Therefore the average $ac$ is 1.

Note also that
\[  b(\mathbf{Co}) = b(\mathbf{Cn}) = b(\mathbf{Ct}) \]
\begin{figure}

\centerline{%
\includegraphics[height=0.96\textwidth,bb=27 111 892 775,angle=90,clip=]{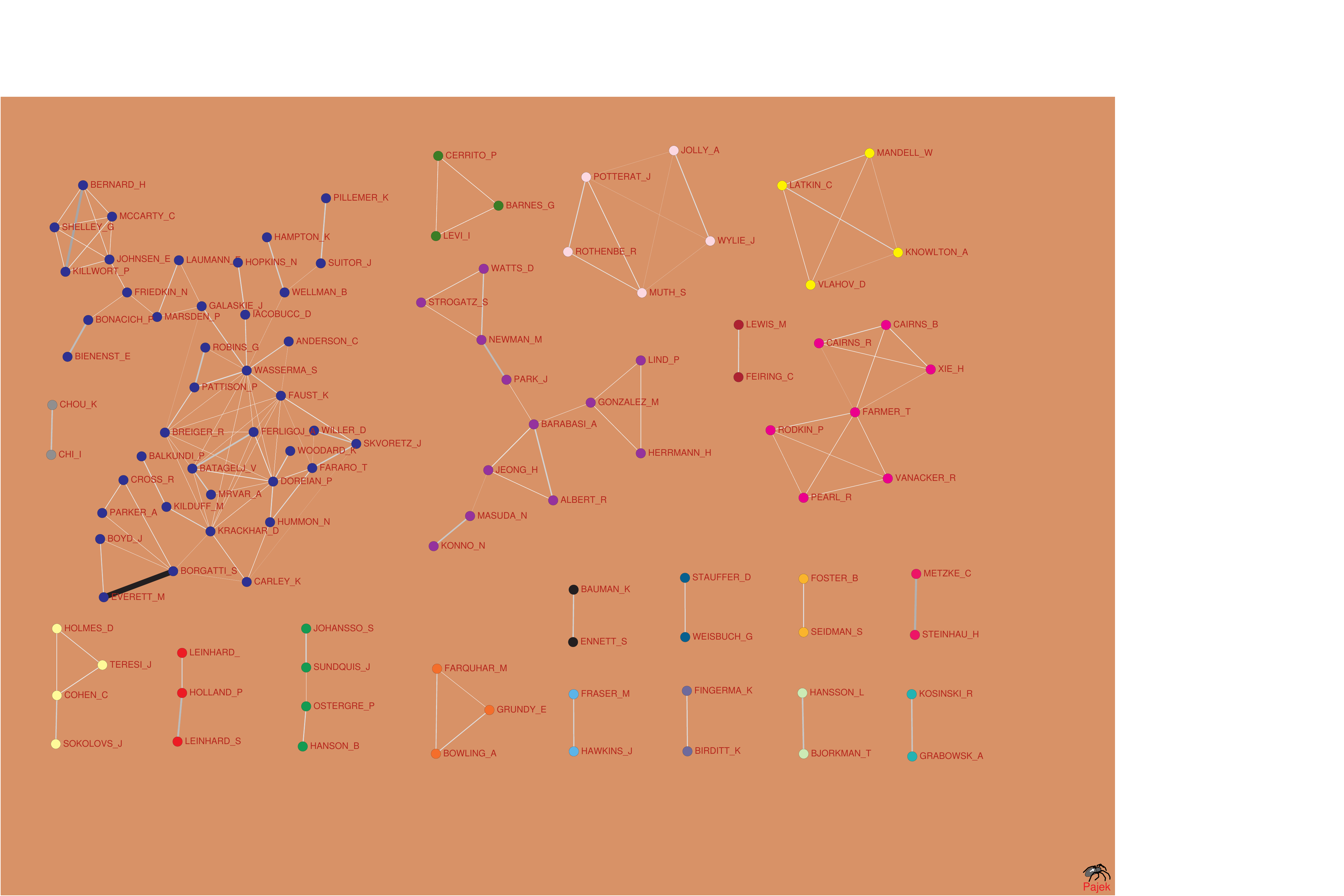}}
\caption{$p_S$-core of order 0.75 in the third collaboration network on
$\mathbf{Ct}$(\texttt{SN5})}\label{psc}

\end{figure}

In Figure~\ref{psc} the $p_S$-core of order 0.75 in the third collaboration network
$\mathbf{Ct}$(\texttt{SN5}) is presented. In this core the large cliques disappear. The largest core's
component consists of the main stream social networks researchers with the most
intensive collaboration pairs: Borgatti and Everett, Killworth and Bernard, Bonacich
and Bienestock, Ferligoj and Batagelj, Pattison and Robins, etc. The second largest
component consists of physicists with more intensive collaboration pairs: Newman and
Park, Barabasi and Albert, and Masuda and Konno. In the smaller components we find
additional three pairs: Lienhard and Holland (with Lienhard represented with two nodes),
Metzke and Steinhaus, and Chou and Chi.

\section{Derived networks}

\subsection{Bibliographic Coupling and Co-Citation}

In \WoSPajek the citation relation means
$ p \,\mathbf{Ci}\, q \equiv p \textrm{~cites~} q $.
Therefore the \keyw{bibliographic coupling} network $\mathbf{biCo}$ can be determined as
\citep{bico}
\[ \mathbf{biCo} = \mathbf{Ci} * \mathbf{Ci}^T \]
The corresponding weight
\[ bico_{pq} = \sum_{s \in W} ci_{ps} ci_{qs} = \sum_{s \in N(p)\cap N(q)} 1 \]
is equal to the number of works cited by both works $p$ and $q$. It is symmetric
$ bico_{pq} = bico_{qp} $.

Again we have problems with works with many citations, especially with review
papers. To neutralize their impact we can introduce a normalized measure such as
\[ \mathbf{biCon} = \frac{1}{2}( n(\mathbf{Ci}) * \mathbf{Ci}^T +
   \mathbf{Ci} * n(\mathbf{Ci})^T ) \]
It is easy to verify that $bicon_{pq} \in [0,1]$ and $bicon_{pq}=bicon_{qp}$
(symmetry). It also holds: $bicon_{pq} = 1$ iff the works $p$ and $q$ are
referencing the same works. Note that
\[ b( n(\mathbf{Ci}) * \mathbf{Ci}^T ) = b(\mathbf{Ci} * n(\mathbf{Ci})^T ) . \]

The $cC_{pq}$ element of the first term represents the 'importance' of common
$(p,q)$-citations for the work $p$; and the $C\!c_{pq}$ element of the second term
represents the 'importance' of common $(p,q)$-citations for the work $q$.
\[  bicon_{pq} = \frac{1}{2}( cC_{pq} + C\!c_{pq}) \]
Note that the first term in the definition of $\mathbf{biCon}$ is equal to the
transpose of the second term
\[ (\mathbf{Ci} * n(\mathbf{Ci})^T ))^T = n(\mathbf{Ci}) * \mathbf{Ci}^T \]
and therefore $C\!c_{pq} = cC_{qp}$.
This can be used for more efficient computation of $\mathbf{biCon}$. We only
need to compute the first term $\mathbf{cC}$. Then
\[  bicon_{pq} = \frac{1}{2}( cC_{pq} + cC_{qp}) \]

In the network $\mathbf{biCon}(\texttt{SN5})$ the larger components with edges with
$bicon = 1$ correspond to papers with a single reference to a book (
Wasserman, S., Faust, K.: Social network analysis. Cambridge UP, 1994;
Taylor, Howard F.: Balance in small groups. Van Nostrand Reinhold, 1970;
Belle, D.: Childrens social networks and social supports. Wiley, 1989;
Gottlieb, B. H.: Social networks and social support. Sage, 1981;
Yan, Yunxiang: The flow of gifts. Stanford UP, 1996;
Zhang, L.: Strangers in the City. Stanford UP, 2001).
There are also several pairs of papers with $bicon = 1$, mostly written by the
same author. More interesting groups we can obtain as larger islands with
values below 1.

Similary the \keyw{document co-citation} network $\mathbf{coCi}$ can be determined as
\citep{ker,coci}
\[ \mathbf{coCi} = \mathbf{Ci}^T * \mathbf{Ci} \]
The corresponding weight
\[ coci_{pq} = \sum_{s \in W} ci_{sp} ci_{sq} = \sum_{s \in N^-(p)\cap N^-(q)} 1 \]
is equal to the number of works citing both works $p$ and $q$. $N^-(p)$ denotes the
set of neighbors from which the node $p$ can be entered.

It holds $\mathbf{coCi}(\network{N}) = \mathbf{biCo}(\network{N}^T)$
and also for corresponding normalized networks
$\mathbf{coCin}(\network{N}) = \mathbf{biCon}(\network{N}^T)$.


\subsection{Other derived networks}

The weight $aci_{ip}$ in the \keyw{author citation} network
\[ \mathbf{ACi} = \AW * \mathbf{Ci} \]
counts the number of times author $i$ cited work $p$.

The \keyw{author co-citation} network can be obtained as
\[ \mathbf{ACo} = b(\mathbf{ACi}) * b(\mathbf{ACi})^T \]
The weight $aco_{ij}$ counts the number of works cited by both authors
$i$ and $j$.

The weight $ak_{ik}$ in the \keyw{authors using keywords} network
\[ \mathbf{AK} = \AW * \mathbf{WK} \]
counts the number of works in which the author $i$ used a keyword $k$.

\subsection{The cited co-authorship network}

\cite{Qua} proposed the \keyw{cited co-authorship network}:
\[ \AW * \diag(\indeg_\mathbf{Ci}(p)) * \WA \]
the weight of two collaborating authors
is equal to the sum of numbers of citations to co-authored works
where $\indeg_\mathbf{Ci}(p)$ is number of citations to work $p$.

Its normalized version would be:
\[ \mathbf{Cc} = \AW * \diag\left(\frac{\indeg_\mathbf{Ci}(p)}{\outdeg_\mathbf{Ci}(p)^2}\right) * \WA \]
with the properties
\[  \sum_{i \in A} \sum_{j \in A} wa_{ip} \frac{\indeg_\mathbf{Ci}(p)}{\outdeg_\mathbf{Ci}(p)^2} aw_{pj} = \indeg_\mathbf{Ci}(p) \]
and
\[  \sum_{i \in A} \sum_{j \in A} cc_{ij} = \sum_{p \in W} \indeg_\mathbf{Ci}(p) = | \edges{A}_\mathbf{Ci} | \]
where $| \edges{A}_\mathbf{Ci} |$ is the number of arcs in the citation network
$\mathbf{Ci}$.

\subsection{Authors' citations network}

\begin{figure}
\centerline{\includegraphics[width=0.8\textwidth,bb=20 5 380 215,clip=]{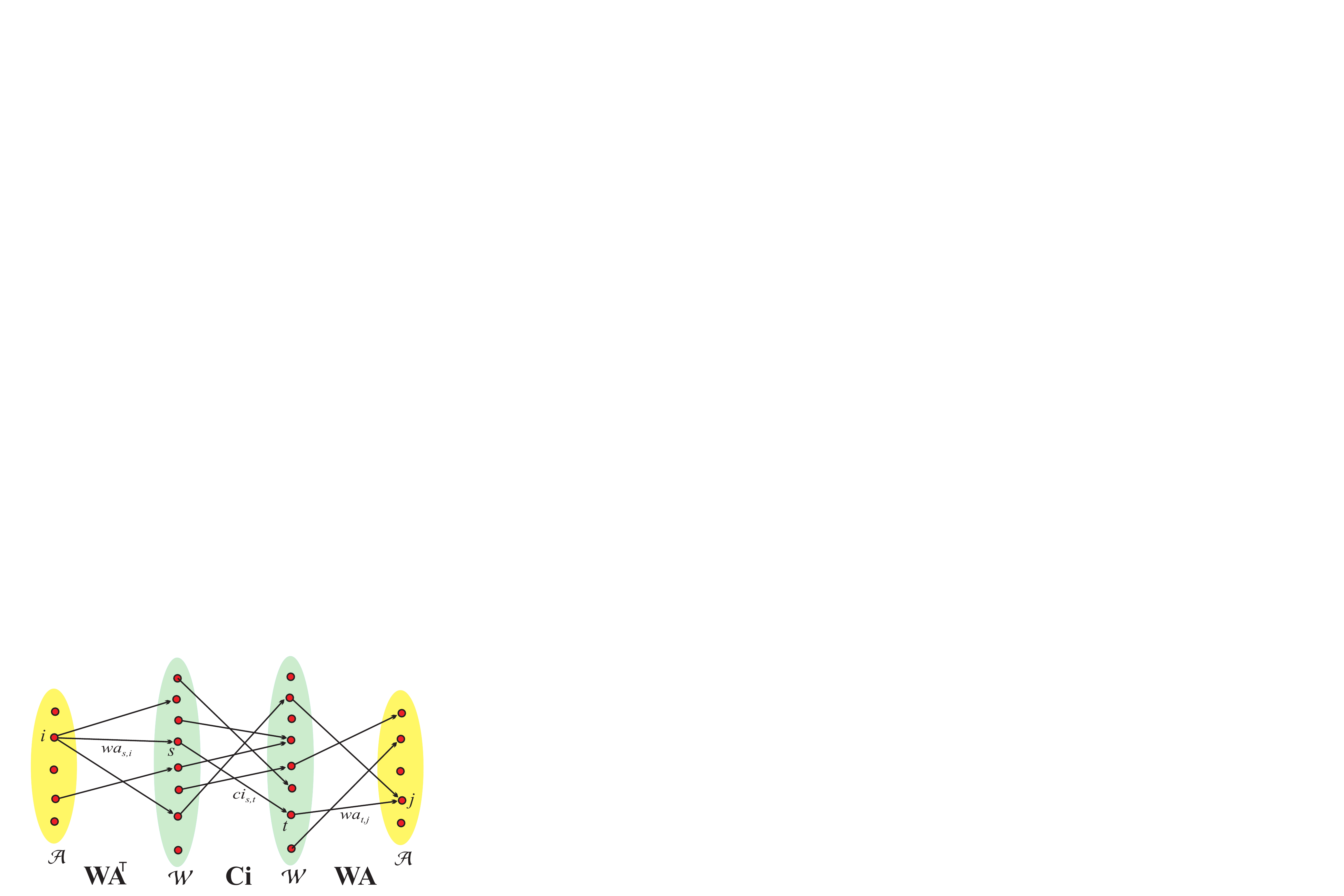}}
\caption{Authors' citations network}\label{WtCW}
\end{figure}

The network of citations between authors can be obtained as
\[ \mathbf{Ca} = \AW * \mathbf{Ci} * \mathbf{\WA} \]
The weight $ca_{ij}$ counts the number of times a work co-authored by
$i$ is citing a work co-authored by $j$, see Figure~\ref{WtCW}.

\begin{figure}
\centerline{\includegraphics[width=0.88\textheight,bb=0 0 1023 738,clip=,angle=90]{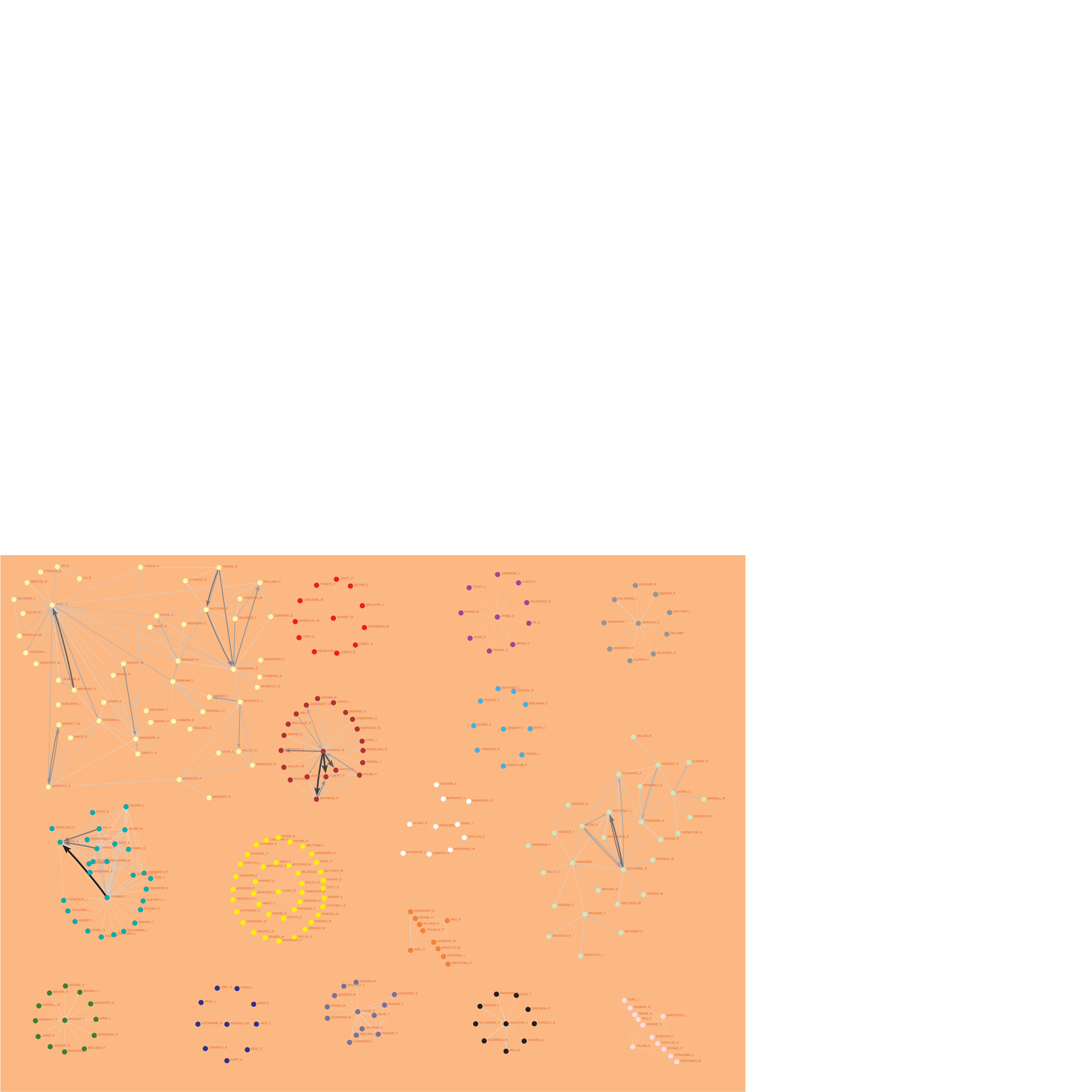}}
\caption{Some link islands in the network $\mathbf{Ca}$(\texttt{SN5})}\label{isla}
\end{figure}

In Figure~\ref{isla} some link islands from $\mathbf{Ca}$(\texttt{SN5}) are
presented. The largest island consists of the main stream social networks
researchers with some subgroups: the star around R. Burt in the top left
part; the S. Borgatti and M. Everett tandem in the bottom left part;
the probabilistic group in the top right part with G. Robins, P. Pattison,
T. Snijders, S. Wasserman, and P. Holland as the most prominent; and others:
J. Skvoretz, D. Krackhardt, P. Doreian, R. Breiger, H. White, L. Freeman,
and P. Marsden.

The "scale-free" island consists mainly of physicists M. Newman, A. Barabasi,
D. Watts, R. Albert, P. Holme and others. In the "medical" island the central
authors are J. Potterat, R. Rothenberg, D. Woodhouse, S. Muth, A. Klovdahl,
and S. Friedman. There is also an island on "education and psychology" with
T. Farmer, R. Cairns, B. Cairns, H. Xie, and P. Rodkin.

Most of the other islands are star-like, usually a professor with his
Phd students.

\section{Conclusions}

In the paper we showed that the bibliographic data can be transformed into a
collection of compatible networks. Using network multiplication different
interesting derived networks can be obtained. In defining them an appropriate
normalization should be considered. The proposed approach can be applied also
to other collections of compatible networks \citep[see][pg. 8260--8262]{Ency}.

Note that most of the obtained derived networks are one-mode networks.
To analyse them standard SNA methods can be used. For analysis of two-mode
networks we can use direct methods such as (generalized) two-mode cores,
two-mode hubs and authorities and 4-rings islands \citep{Ahmed}.

We can also transform the citation network (and other WoS networks) into temporal
network using the partition of works by publication year.
Using the time slices also the temporal sequences of corresponding derived networks
can be obtained.

\Pajek allows analyses on different levels specified by a partition of the
corresponding set of units and obtained using  the \keyw{shrinking} of classes.
For example: partition of authors by institutions, or partition of institutions
by countries, partitions of authors by discipline/ field/
subfield, etc. Using the \keyw{extraction} of selected classes we can reduce
the network to the area of our interest.

The HOW TO in \Pajek for the described approach is available at \\
\texttt{http://pajek.imfm.si/doku.php?id=how\_to:biblio}

\section{Acknowledgments}

The work was supported in part by the ARRS, Slovenia, grant P1-0294, as well as by
grant within the EUROCORES Programme EUROGIGA (project GReGAS) of the European
Science Foundation. The second author was financed in part by the European Union, European Social Fund.

\appendix
\section{References for analyses}\label{appendix}
\renewcommand{\refname}{}

\renewcommand{\refname}{References}

\end{document}